# Constitutive modeling of the densification process in silica glass under hydrostatic compression

V. Keryvin[*,a,b], J.-X. Meng[a,b], S. Gicquel[b,1], J.-P. Guin[b], L. Charleux[b,c], J.-C. Sanglebœuf[b], P. Pilvin[a], T. Rouxel[b], G. Le Quilliec[b,2]

[a]Univ. Bretagne-Sud, EA 4250, LIMATB, F-56321 Lorient, France
[b]Univ. Rennes 1, ERL CNRS 6274, LARMAUR, F-35042 Rennes, France
[c]Univ. Savoie, EA 4144, SYMME, F-74940, Annecy-le-Vieux, France

**Abstract**

The mechanical response of amorphous silica (or silica glass) under hydrostatic compression for very high pressures up to 25 GPa is modelled via an elastic-plastic constitutive equation (continuum mechanics framework). The material parameters appearing in the theory have been estimated from the *ex situ* experimental data from Rouxel *et al.* [Rouxel T, Ji H, Guin JP, Augereau F, Rufflé B J Appl Phys 2010;107(9):094903]. The model is shown to capture the major features of the pressure-volume changes response from the *in situ* experimental work of Sato and Funamori [Sato T, Funamori N Phys Rev Lett 2008;101:255502]. In particular, the onset and saturation of densification, the increase in elasticity parameters (bulk, shear and Young's moduli) and Poisson's ratio are found to be key parameters of the model.

[*]Corresponding author: vincent.keryvin@univ-ubs.fr
[1]Present address: EDF – DPI, DCN/Pôle Assemblage Combustible; 1, place Pleyel 93282 Saint-Denis Cedex, France.
[2]Present address: Laboratoire Roberval, UMR 7337 UTC-CNRS, Université de Technolo-

## 1. Introduction

Because of their relatively low atomic packing density compared to their crystalline counterparts, glasses experience significant densification (permanent increase in density) under high hydrostatic pressures. As a matter of fact, the density of a-SiO$_2$ (amorphous silica) can be increased by up to 20 % and that of window glass by 6 %, when a sufficiently high hydrostatic pressure is applied [1, 2, 3, 4, 5, 6].

Permanent modifications in silica glasses density is difficult to investigate *via* unconstrained macroscopical testing (such as the compression test) because of the material brittleness. On the contrary, hydrostatic compression on small volumes of material impede drastically cracking: permanent strains can be observed without any cracking features when possible spurious effects of additional shear are absent [7, 8]. These tests usually give, after unloading (*ex situ*), some information on the density changes. The combination of such tests with physical spectroscopy techniques (X-Ray Diffraction, Raman scattering, Brillouin scattering), *e.g.* in a diamond-anvil cell, permits to follow *in situ* the changes in the structure of silica glass (short-to-medium range order). However, from a mechanical point of view, the *in situ* mechanical response of the test is partial as only the pressure information is known, not the density one, during the test[1].

Recent advances in experimental testing have made it possible to obtain

---

gie de Compiègne, France.

[1]It is however possible to extract this information assuming a purely elastic behaviour for pressures lower than the densification threshold [7].

the *in situ* mechanical response of the hydrostatic compression test (curve pressure-volume changes). Sato and Funamori [9] conducted experiments up to 60 GPa with a diamond-anvil cell at room temperature. The density of the silica glass sample was determined *in situ* from the intensities of transmitted X-rays measured for the sample and some reference materials (see Figure 1). Apart from this mechanical response, they related their experiments to structure changes by using X-Ray absorption and diffraction techniques [9, 10]. They found that silica glass behaves as a single amorphous polymorph having a fourfold- coordinated structure below 10 GPa. Irreversible changes in the intermediate-range order begin at around 10 GPa (referred to as densification), up to 25 GPa. It corresponds to an irreversible and progressive transformation from a low density amorphous phase to a high density amorphous phase. This latter phase is characterised by an increase in the statistic distribution of 4- and 3-membered rings of $SiO_4$ tetrahedra with a narrowing of inter-tetrahedral angle distribution [11, 12].

From a more mechanistic point of view, the deformation mechanisms, between 0 and 25 GPa, may be depicted as follows [1, 2, 3, 4, 6, 13]. Below a threshold pressure, the behaviour is purely elastic. Above a second threshold pressure, further referred to as saturation pressure, the behaviour is once again purely elastic. In between these two pressures, densification occurs and develops by increasing the applied pressure (referred to as hardening[2])

---

[2]Strain hardening is commonly ascribed to volume conservative plasticity where it is classically defined as the increase in flow stress upon plastic flow. Under such circumstances, the maximum strain is limited by the material strength. Otherwise, in the case of perfect

and the elastic moduli increase with the densification level.

Such a summary of structural changes, deformation mechanisms and constitutive models is found in Figure 2.

Prior to this recent advance in experimental testing, the modelling of permanent deformation in glasses has been based on constrained mechanical tests that make it possible to develop stable permanent deformation fields without fracture or even cracking. This is, for instance, the case during hardness or scratch experiments. For temperatures well below the glass transition, according to the literature, the formation of the residual imprint is thought to result from the concomitant contribution of two deformation mechanisms: densification and shear flow [14, 15, 16, 17, 18, 5, 19]. Constitutive models were developed to clarify this issue on the hardness of glass [14, 20, 21, 22, 23]. They may involve only volume-conservative plasticity (further referred to as plasticity) – therefore being unable to predict densification! [14]–, densification and plasticity [20, 21, 22] and even hardening [23]. The two latter models are based on the correct description of the instrumented indentation test response. Instrumented indentation test enriches the hardness test by giving access to the load vs. penetration curve. These new data are used to suggest more realistic constitutive equations in

---

plasticity, strains over 1 (superplasticity) could be possibly achieved because there is no geometrical constraint to shear processes. On the contrary, densification is a geometrically constrained process, where pressure can be ideally increased to infinity without fracture, while strain is limited by the details of the atomic packing characteristics. It is obvious that densification becomes more and more difficult as the density increases. This has nothing to do with strain- or time-hardening processes observed in metal plasticity. Nevertheless, we will use the term hardening in this text.

a straightforward way. The indentation test is heterogeneous by nature and as a consequence, numerical simulations by the finite-element method are generally used to link given material properties to the load vs. penetration curve and the residual imprint. Material parameters are then estimated using a identification procedure. Such modelings have been proposed these last fifteen years, notably in the key works of Lambropoulos [22, 24] and Kermouche *et al.* [23]. Both models assume that a combination of pressure and shear terms triggers permanent deformation (densification and plasticity). In all these models, attention has been paid mainly on the role of shear on the permanent deformation process. It appears, from the survey of these constitutive equations and the numerical simulations made with them, that different models allow one to fit the load-displacement curve in instrumented indentation [22, 23]. It is noticed also, that some models do not consider the hardening-like behaviour of densification [20, 22]. Moreover, no models takes into account the saturation in densification as well as the changes in elastic parameters with densification. Meanwhile, not much effort has been devoted to what takes place on the hydrostatic axis, that is during hydrostatic compression.

The purpose of this paper is therefore to focus on the sole densification process under very high hydrostatic pressures, with a new constitutive model allowing one to take into account the onset of densification above a threshold pressure, its increase with applied pressure, its saturation above a second threshold pressure, and the changes in elastic parameters coupled with the densification process. We restrict the applied pressures to levels below 25 GPa to avoid other deformation processes not linked to the sole densification. The choice is made to deal with a continuum mechanics framework rather than with discrete mesoscopic models, such as molecular dynamics (see *e.g.* [25, 26]), to allow, in the near future, the simulation of complicated processes, such as indentation or scratching, requiring large numbers of atoms and large simulation times, still unreachable so far with mesoscopic models. Yet, these discrete modelings make it possible to extract valuable information on the short-to-medium range order in glasses, which can be compared to precise experiments with physical spectroscopy techniques. Eventually, an interesting and promising way is multi scale modelling [27]. The paper is organised as follows. The new constitutive model is outlined in Section 2. In Section 3, we use *ex situ* experimental data for silica glass [19] to estimate the material parameters appearing in our specialised constitutive equations. We have implemented our constitutive model in the finite element program ABAQUS/Standard [28], and we report on a finite-element simulation of the hydrostatic compression test, in Section 4. We compare the simulation results to *in situ* experimental data [9, 29]. We close in Section 5 with some final remarks.

## 2. Constitutive model

### 2.1. General considerations

We limit our considerations to isothermal situations at room temperature. The material is assumed to be isotropic. The material point in the reference configuration and in the deformed configuration are denoted respectively by $\underline{X}$ and $\underline{x}$. We use the standard notation of modern continuum mechanics[3]: $\underline{x} = \underline{x}(\underline{X}, t)$, motion; $\underset{\sim}{F} = \dfrac{\partial \underline{x}}{\partial \underline{X}}$, deformation gradient; $J = \det \underset{\sim}{F}$, its jacobian; $\underset{\sim}{\sigma}$, Cauchy stress tensor.

To fulfill the material frame indifference, we use a local objective frame, the corotational frame in our case (see the review of Xia *et al.* [30]), where the constitutive equations are written in a similar way as in the small-strain case. This local frame is transformed vis-à-vis the initial frame, by a rotation. In this frame, we decompose the rate of deformation, $\underset{\sim}{\dot{e}}$, into the sum of an elastic part, $\underset{\sim}{\dot{e}}^e$, and a plastic part, $\underset{\sim}{\dot{e}}^p$.

This rate approach naturally introduces a logarithmic measure of strain, taking into account the large strains (Hencky strain tensor), developed during densification.

## 2.2. Application to the hydrostatic loading case

We limit our attention to the hydrostatic loading case, where the motion is $\underline{x} = \underline{X} + \lambda \underline{X}$, for $\lambda \leq 0$ ($\lambda$ is the relative change in length). The transformation gradient $\underset{\sim}{F}$ is $(1+\lambda)\underset{\sim}{i}$ ($\underset{\sim}{i}$ is the second order unit strain tensor). The jacobian is $J = (1+\lambda)^3$. The Green–Lagrange strain tensor, $\underset{\sim}{E} = \frac{1}{2}(\underset{\sim}{F}^T \cdot \underset{\sim}{F} - \underset{\sim}{i})$, is $((1+\lambda)^2 - 1)\underset{\sim}{i}$. The small strain tensor, $\underset{\sim}{\epsilon}$, is $\lambda \underset{\sim}{i}$. The mass conservation in its lagrangean form gives $\rho_0 = \rho J$, so that $J = \dfrac{\rho_0}{\rho} = \dfrac{V}{V_0} = (1+\lambda)^3$.

For the small strains assumption, $|\lambda| \leq 1$, we have $\dfrac{V}{V_0} = (1+\lambda)^3 \simeq 1 +$

---
[3]$\underline{a}$ (resp. $\underset{\sim}{a}$) denotes a vector field (resp. a second order tensor field)

$3\lambda = 1 + \operatorname{tr}\underset{\sim}{\epsilon}$, so that, the volumetric strain, $\varepsilon_v^{\mathsf{SS}}$, is:

$$\varepsilon_v^{\mathsf{SS}} = \operatorname{tr}\underset{\sim}{\epsilon} = 3\lambda = \frac{V}{V_0} - 1 = \frac{\Delta V}{V_0} = J - 1 \qquad (1)$$

As for finite strains, the Hencky strain tensor, which is the strain measure, is $\underset{\sim}{h} = \frac{1}{2}\ln(\underset{\sim}{F} \cdot \underset{\sim}{F}^T)$, and, in the case of hydrostatic compression, is $\ln(1+\lambda)\underset{\sim}{i}$. Thus, the volumetric strain, $\varepsilon_v^{\mathsf{FS}}$, is:

$$\varepsilon_v^{\mathsf{FS}} = \operatorname{tr}\underset{\sim}{h} = 3\ln(1+\lambda) = \ln\frac{V}{V_0} = \ln J = \ln\frac{\rho_0}{\rho} = \ln(1+\varepsilon_v^{\mathsf{FS}}) \qquad (2)$$

Since, there is no rotation between the corotational frame and the reference one, for this very loading case, the strain $\underset{\sim}{e}$ in the corotational frame is the Hencky strain tensor $\underset{\sim}{h}$.

*2.3. Permanent volume and density changes*

We define $\alpha$, the densification level as:

$$\alpha = \frac{\Delta\rho^p}{\rho_0} = \frac{\rho^p - \rho_0}{\rho_0} \geq 0 \qquad (3)$$

where $\rho_0$ is the initial density and $\Delta\rho^p$ is the difference between the actual density after unloading, $\rho^p$, and $\rho_0$.

We decompose the volumetric strain, $\varepsilon_v$, into the sum of an elastic part, $\varepsilon_v^e$, and a plastic part, $\varepsilon_v^p$. As recalled, the mass conservation allows one to link density changes and volume changes, so that the small strains plastic volumetric strain, $\varepsilon_v^{SS,p}$ reads:

$$\varepsilon_v^{SS,p} = -\frac{\alpha}{1+\alpha} \leq 0 \tag{4}$$

while the finite strains plastic volumetric strain, $\varepsilon_v^{FS,p}$ reads (using Eq. (2)):

$$\varepsilon_v^{FS,p} = \ln(1 + \varepsilon^{SS,p}) = -\ln(1+\alpha) \leq 0 \tag{5}$$

### 2.4. Model features

The set of constitutive equations (elastic behaviour, yield function, flow rule, changes in elastic moduli) is summarised below within a viscoplastic scheme. It is written in the corotational frame. For sake of clarity the superscript FS is omitted in the following for the volumetric strain $\varepsilon_v$.

(i) Linear elastic behaviour

By using Hooke's law, we have a linear relationship between pressure and the elastic part of the volumetric strain:

$$P = -B\,\varepsilon_v^e \tag{6}$$

where $P$ is the hydrostatic pressure ($P = -1/3\,\mathrm{tr}\,\underset{\sim}{\sigma}$) and $B$ is the bulk modulus. We assume the elastic behaviour to be and to remain linear. The bulk modulus may nevertheless change during the deformation process.

(ii) Yield function: onset of densification, hardening and saturation

The yield function is used for knowing whether the behaviour is elastic ($f \leq 0$) or elasto-viscoplastic ($f > 0$). We write:

$$f(P; P_a) = P - P_a(\varepsilon_v^p) \tag{7}$$

where $P_a$ is the actual value of the pressure threshold, below which no densification occurs (when densification is not saturated). $P_a(\varepsilon_v^p = 0) = P_0$ and $P_a(\varepsilon_v^p = \gamma) = P_1$ are the onset and saturation pressures respectively; $\gamma$ is the saturation value of densification. For $P < P_a$ or $P > P_1$, the behaviour is purely elastic.

(iii) Viscoplastic flow rule

Assuming an associative flow rule (the plastic part of the strain rate is normal to the yield surface $f = 0$), we have

$$\dot{\underset{\sim}{e}}^p = \left\langle \frac{f}{K} \right\rangle^n \frac{\partial f}{\partial \underset{\sim}{\sigma}} = -\frac{1}{3} \left\langle \frac{f}{K} \right\rangle^n \underset{\sim}{i} \quad \text{for } P \leq P_1, \quad 0 \text{ for } P > P_1 \tag{8}$$

where $n$ and $K$ are viscoplastic parameters of this power-like law [31][4].
And the volumetric strain rate is:

$$\dot{\varepsilon}_v^p = - \left\langle \frac{f}{K} \right\rangle^n \quad \text{for } P \leq P_1, \quad 0 \text{ for } P > P_1 \tag{9}$$

For $P \leq P_1$ (densification not saturated), $\dot{\varepsilon}_v^p = 0$ when $f \leq 0$ and $\dot{\varepsilon}_v^p > 0$

---
[4] $<x>$ denotes the positive part of $x$.

when $f > 0$. When the saturation level of densification is reached ($\varepsilon_v^p = \gamma$), the deformation process becomes again purely elastic ($\dot{\varepsilon}_v^p = 0$).

(iv) Changes in elastic moduli: elasticity-densification coupling

The changes in the elastic moduli $M$ (bulk and shear moduli[5]) are described by:

$$M(\alpha) = M_0 + (M_{\max} - M_0) \times \mathscr{C}(\frac{\varepsilon_v^p}{\gamma}) \qquad (10)$$

where $\mathscr{C}$ is a function so that $\mathscr{C}(0) = 0$ and $\mathscr{C}(1) = 1$. This function may differ from one modulus to another one. $M_0$ and $M_{\max}$ denote the initial value of the considered modulus and its saturated value, respectively. We do not consider in this study the possibility of reversible changes in elastic moduli due to the very high pressures (third order elastic constants, see *e.g.* [32, 33]). Thus, the value of a modulus $M$ at a given applied pressure, therefore at a given value of the densification level, has the same value at ambient pressure, after unloading.

## 3. Estimates of material parameters for silica glass from *ex situ* experiments

In this Section, we refer to *ex situ* experimental data from Rouxel *et al.* [19] to estimate material parameters appearing in the preceding constitutive equations.

---

[5]Taking into account the changes in the shear modulus has no influence on the following numerical simulations, since only the bulk modulus plays a role in a pure densification process. However, we keep it for a more general perspective.

These references allow one to extract the permanent changes in density, $\alpha$, after unloading, for different levels of applied pressures. They are represented in Figure 3. The saturation level of densification is estimated as 21.6%. We set, using Eq. (5),

$$\gamma = -\ln(1+0.216) \simeq -0.196$$

The evolution of densification as a function of applied pressure may be represented by using a fitting function to deal with these progressive changes. It is inspired by the Johnson-Mehl-Avrami-Kolmogorov (JMAK) function used for the nucleation and growth of crystals in crystallisation kinetics (see for instance Ref. [34]). Let $Y = \frac{\varepsilon_v^p}{\gamma}$ and $x = \frac{P - P_0}{P' - P_0}$, the JMAK equation writes:

$$Y(x) = \left(1 - \exp(-k \times x^m)\right) \times H(x) \quad (11)$$

where $H$ is the step function and $k$, $m$ and $P'$ are fitting parameters. A simple hand-made fit gives satisfactory results as seen in Figure 3. We have:

$$P_0 = 3 \text{ GPa} \quad ; \quad P' = 20 \text{ GPa} \quad ; \quad P_1 = 25 \text{ GPa} \quad ; \quad m = 4 \quad ; \quad k = 3$$

Contrary to $P'$, which is slightly lower than $P_1$, $P_0$ keeps its physical meaning as onset of densification.

From Ref. [19], we also extract the changes in elastic parameters with the densification level, namely bulk and shear moduli (see Figure 4). We evaluate the ambient pressure initial ($B_0$, $G_0$) and saturation parameters, after unloading of a 25 GPa hydrostatic compression ($B_{max}$, $G_{max}$), for the bulk and shear moduli. We have:

$$B_0 = 35.5 \text{ GPa} \quad ; \quad G_0 = 32.5 \text{ GPa} \quad ; \quad B_{max} = 73.3 \text{ GPa} \quad ; \quad G_{max} = 43.1 \text{ GPa}$$

We observe that the coupling between elasticity and densification is strong since the bulk modulus will more than double! At a lesser extent, the shear modulus will also increase, but moderately. As a consequence, Poisson's ratio will reach values around 0.25 (the initial value is 0.15) as shown in Figure 5. The $\mathscr{C}$ functions defined in Eq. (10) were taken linear as represented in Figure 4. For the shear modulus, this seems to be a good approximation. As for the bulk modulus, it gives an overestimated approximation although the lack of information between 4% and 20% of densification prevents from using a more relevant one. This is a way for improvement.

The viscoplastic parameters, $n$ and $K$, have been set to give a rate-independent response, thus neglecting, in this paper, any effects of time and loading rate. We have:

$$n = 2 \quad ; \quad K = 1 \text{ MPa} \cdot \text{s}^{\frac{1}{n}}$$

All parameter values are presented in Table 1.

# 4. Finite-element simulation of the hydrostatic compression test and comparison to *in situ* experiments

We have implemented our constitutive model in the finite-element computer program ABAQUS/Standard [28] by writing a user material subroutine. The set of equations is, in a standard way, written in the corotational framework in the subroutine (see Section 2). The mechanical fields (stress, strain) are then calculated by the computer program in the deformed framework [28, 30] Numerical tests have been performed to mimic different models available in the computer program library to ensure that no spurious effects were triggered by using this procedure. The hydrostatic compression test has been simulated under prescribed applied pressures for values up to 25 GPa, above which phase transformations are likely to occur and densification is saturated [35, 9], then down to zero. The mesh consists in only one linear three-dimensional finite element as the test is homogeneous. We use the computer program to deal with the geometrical non linearities of the problem (finite strains).

In the aim of observing the impact of taking into account the fine material behaviour (as reported in Section 1), the constitutive models of Lambropoulos [22] and Kermouche *et al.* [23] have also been implemented in the computer program, with the material parameters reported in their respective manuscripts, using the procedure depicted in Section 2. To make sure that each implementation was correctly made, we have lead a numerical simulation of the indentation test (axisymmetric conditions, frictionless contact).

The results were the same as those reported in their papers with threshold pressures of 9 and 11.5 GPa, respectively.

The *ex situ* information extracted from Ref. [19] provides quantitative values of the densification behaviour as well as on the changes in elastic parameters. We wonder whether our constitutive model detailed in Section 2, along with the material parameters extracted in Section 3, is able to describe the *in situ* mechanical response of the test, *i.e.* the pressure-density curve taken from Ref. [9] for the loading part on pristine samples compressed up to full densification and Ref. [29] for the unloading one on fully densified samples compressed in a reversible way (Figure 1). The first paper [9] allows one, to extract the pressure-volume changes or pressure-density changes at room temperature up to 25 GPa without appreciable irreversible changes in the short-range order. The density changes from 2.2 g/cm$^3$ (initial density $\rho_0$) at zero pressure to $\sim$ 3.6 g/cm$^3$ at $\sim$ 23 GPa. We combine this with data on a fully densified silica glass sample (initial density of 2.67 g/cm$^3$) compressed below 10 GPa [29]. Figure 6 is a modification of Figure 1, where the x-axis stands for the volumetric changes, calculated using Eq. (2).

The results from numerical simulations accounting or not for the changes in elastic moduli, along with the models of Lambropoulos (label 1) and Kermouche *et al.*(label 2), are plotted in Figure 6. For a given applied pressure of 25 GPa, the model from Kermouche *et al.* (linear hardening) predicts volumetric changes $\sim$ -1.2. The model developed in this paper predicts a value of $\sim$ -0.9 when accounting only for the saturation in densification (la-bel 3). When taking into account, in addition, the changes in elastic moduli the value is ∼ -0.5. As for the model from Lambropoulos, it is given only for comparison, since the pressure is limited to the yield pressure of 9 GPa (no hardening). Considering or not both the saturation in densification and the changes in elastic moduli in the constitutive modelling has therefore a tremendous effect on the numerical simulations.

Moreover, the comparison is made in Figure **??** between experimental data and the complete model is very good. It must be emphasised that, at low pressures (2-5 GPa), the model does not take into account the elastic softening, which is a reversible process and corresponds to a decrease of the Si-O-Si inter-tetrahedral angle [7, 12]. At for the high pressures (> 15 GPa), the model underestimates slightly the experimental data. The comparison could be improved when taking into account a non linear change in bulk modulus with densification (see Fig. 4) or by performing a identification of parameters for a very fine tuning. Eventually, the use of a JMAK equation for describing the increase in densification with applied pressure may seem inappropriate since the data collected from Ref. [6] is limited. However, very recently, Deschamps *et al.* [36] proposed a calibration curve linking selective Raman parameters to the degree of densification, by Raman scattering spectroscopy. Data from this paper are added to Figure 3, where a very close correlation to the simple JMAK fit is found.

## 5. Concluding remarks

We have developed a constitutive framework to model the response of silica glass under hydrostatic compression with very high pressures up to 25 GPa. This model takes into account the densification process that exists in between two threshold pressures, the onset of densification and its saturation. It considers the progressive densification with pressure as well as the changes in elastic parameters with densification. The material parameters involved in this model have been determined in a straightforward way from *ex situ* experimental data from Rouxel *et al.* [19].

The model has been implemented in a finite-element software to simulate the hydrostatic compression. Comparisons have been made with other constitutive models, built on hardness or instrumented indentation experiments. The study of the influence of considering both the saturation in densification ($\sim$ 21 %) and the changes in elastic moduli (the ambient pressure bulk modulus will double) have been shown to be considerable with respect to the latter models, which do not take into account these mechanisms. The numerical simulations have been compared to recent *in situ* experimental data from Sato and Funamori [9] and Wakabayashi *et al.* [29] and an excellent agreement was found.

This constitutive framework, restricted to the sole pressure, is expected to pave the way for future developments where the role of pressure is evident but not unique. The impact of an adequate modelling of the densification process triggered by pure pressure, as it is proposed in this work, on con-strained deformation modes, such as surface damage, is a next natural step. This is a key issue for the use in service of glass products, which is yet to be studied.

**Acknowledgements**

Financial support was provided by the Brittany region (PhD grant, ARED 5382 CARAMEL), the French Ministery for Higher Education and Research (PhD grant) and the European University of Brittany (EPT COMPDYNVERRE). They are gratefully acknowledged.

**References**

[1] P. W. Bridgman, I. Šimon, Effects of very high pressures on glass, J. Appl. Phys. 24 (4) (1953) 405–413. 2, 3

[2] H. M. Cohen, R. Roy, Effects of ultrahigh pressures on glass, J. Am. Ceram. Soc. 44 (10) (1962) 523–524. 2, 3

[3] E. B. Christiansen, S. S. Kistler, W. B. Gogarty, Irreversible compressibility of silica glass as a means of determining the distribution of force in high-pressure cells, J. Am. Ceram. Soc. 45 (4) (1962) 172–177. 2, 3

[4] J. D. Mackenzie, High-pressure effects on oxide glasses: I, Densification in rigid state, J. Am. Ceram. Soc. 46 (10) (1963) 461–470. 2, 3

[5] H. Ji, V. Keryvin, T. Rouxel, T. Hammouda, Densification of window glass under very high pressure and its relevance to Vickers indentation, Scripta Mater. 55 (12) (2006) 1159–1162. 2, 4

[6] T. Rouxel, H. Ji, T. Hammouda, A. Moreac, Poisson's ratio and the densification of glass under high pressure, Phys. Rev. Lett. 100 (225501) (2008) 225501. 2, 3, 16

[7] C. S. Zha, R. J. Hemley, H. K. Mao, T. S. Duffy, C. Meade, Acoustic velocities and refractive index of SiO$_2$ glass to 57.5 GPa by Brillouin scattering, Phys. Rev. B 50 (18) (1994) 13105–12. 2, 16

[8] A. Polian, M. Grimsditch, Sound velocities and refractive-index of densified alpha-SiO$_2$ to 25 GPa, Phys. Rev. B 47 (21) (1993) 13979–13982. 2

[9] T. Sato, N. Funamori, Sixfold-Coordinated Amorphous Polymorph of SiO$_2$ under High Pressure, Phys. Rev. Lett. 101 (25) (2008) 255502. 3, 6, 14, 15, 17, 23, 28

[10] T. Sato, N. Funamori, Reply to Comment on "Sixfold-Coordinated Amorphous Polymorph of SiO$_2$ under High Pressure", Phys. Rev. Lett. 102 (20) (2009) 209604. 3

[11] B. Hehlen, Inter-tetrahedra bond angle of permanently densified silicas extracted from their Raman spectra, J. Phys.: Cond. Matter 22 (2) (2010) 025401. 3

[12] C. Sonneville, A. Mermet, B. Champagnon, C. Martinet, J. Margueritat, D. de Ligny, T. Deschamps, F. Balima, Progressive transformations of silica glass upon densification, J. Chem. Phys. 137 (12) (2012) 124505. 3, 16

[13] A. Perriot, D. Vandembroucq, E. Barthel, V. Martinez, L. Grosvalet, C. Martinet, B. Champagnon, Raman microspectroscopic characterization of amorphous silica plastic behavior, J. Am. Ceram. Soc. 89 (2) (2006) 596–601. 3

[14] D. M. Marsh, Plastic flow in glass, Proc. Royal Soc. London A 279 (1964) 420–435. 4

[15] J. E. Neely, J. D. Mackenzie, Hardness and low-temperature deformation of silica glass, J. Mater. Sci. 3 (6) (1968) 603–609. 4

[16] F. M. Ersnberger, Role of densification in deformation of glasses under point loading, J. Am. Ceram. Soc. 51 (10) (1968) 545–547. 4

[17] K. W. Peter, Densification and flow phenomena of glass in indentation experiments, J. Non-Cryst. Solids 5 (1970) 103–115. 4

[18] S. Yoshida, J.-C. Sanglebœuf, T. Rouxel, Quantitative evaluation of indentation-induced densification in glass, J. Mater. Res. 20 (12) (2005) 3404–3412. 4

[19] T. Rouxel, H. Ji, J. P. Guin, F. Augereau, B. Ruffle, Indentation deformation mechanism in glass: Densification versus shear flow, J. Appl. Phys. 107 (9) (2010) 094903. 4, 6,

11, 12, 15, 17, 25, 26, 27

[20] M. Imaoka, I. Yasui, Finite element analysis of indentation on glass, J. Non-Cryst. Solids 22 (2) (1976) 315–329. 4, 5

[21] I. Yasui, M. Imaoka, Finite element analysis of identation on glass(ii), J. Non-Cryst. Solids 50 (2) (1982) 219–232. 4

[22] J. C. Lambropoulos, S. Xu, T. Fang, Constitutive law for the densification of fused silica, with applications in polishing and microgrinding, J. Am. Ceram. Soc. 79 (6) (1997) 1441–1452. 4, 5, 14

[23] G. Kermouche, E. Barthel, D. Vandembroucq, P. Dubujet, Mechanical modelling of indentation-induced densification in amorphous silica, Acta Mater. 56 (2008) 3222–3228. 4, 5, 14

[24] K. Xin, J. C. Lambropoulos, Densification of fused silica: effects on nanoindentation, in: Proc. SPIE, Vol. 4102, 2000, pp. 112–121. 5

[25] W. Jin, R. K. Kalia, P. Vashishta, J. P. Rino, Structural transformation, intermediate-range order, and dynamical behavior of $SiO_2$ glass at high pressures, Phys. Rev. Lett. 71 (1993) 3146–3149. 6

[26] R. G. Della Valle, E. Venuti, High-pressure densification of silica glass: A molecular-dynamics study, Phys. Rev. B 54 (6) (1996) 3809–3816. 6

[27] D. Rodney, A. Tanguy, D. Vandembroucq, Modeling the mechanics of amorphous solids at different length scale and time scale, Model. Simul. Mater. Sci. Eng. 19 (8) (2011) 083001. 6

[28] Abaqus Analysis User's manual, Version 6.9. Dassault Systemes Simulia Corp., 2010. 6, 14

[29] D. Wakabayashi, N. Funamori, T. Sato, T. Taniguchi, Compression behavior of densified $SiO_2$ glass, Phys. Rev. B 84 (14) (2011) 144103. 6, 15, 17, 23, 28

[30] H. Xiao, O. Bruhns, A. Meyers, Elastoplasticity beyond small deformations, Acta Mech. 182 (1-2) (2006) 31–111. 7, 14

[31] P. Perzyna, Fundamentals problems in viscoplasticity, Adv. Appl. Mech. 9 (1966) 243–

377. 10

[32] D. S. Hughes, J. L. Kelly, 2nd-order elastic deformation of solids, Phys. Rev. 92 (5) (1953) 1145–1149. 11

[33] D. Cavaille, C. Levelut, R. Vialla, R. Vacher, E. Le Bourhis, Third-order elastic constants determination in soda-lime-silica glass by Brillouin scattering, J. Non-Cryst. Solids 260 (3) (1999) 235–241. 11

[34] V. Keryvin, M.-L. Vaillant, T. Rouxel, M. Huger, T. Gloriant, Y. Kawamura, Thermal stability and crystallisation of a $Zr_{55}Cu_{30}Al_{10}Ni_5$ bulk metallic glass studied by an in-situ ultrasonic echography technique, Intermetallics 10 (11-12) (2002) 1289–1296. 12

[35] T. Rouxel, H. Ji, V. Keryvin, T. Hammouda, S. Yoshida, Poisson's Ratio and the Glass Network Topology - Relevance to High Pressure Densification and Indentation Behavior, Adv. Mater. Res. 39-40 (2008) 137–146. 14

[36] T. Deschamps, A. Kassir-Bodon, C. Sonneville, J. Margueritat, C. Martinet, D. de Ligny, A. Mermet, B. Champagnon, Permanent densification of compressed silica glass: a Raman-density calibration curve, J. Phys.: Cond. Matter 25 (2) (2013) 025402. 16, 25

| $B_0$ (GPa) | $B_{max}$ (GPa) | $G_0$ (GPa) | $G_{max}$ (GPa) |
|---|---|---|---|
| 35.5 | 73.3 | 32.5 | 43.1 |

| $P_0$ (GPa) | P' (GPa) | $P_1$ (GPa) | $\gamma$ |
|---|---|---|---|
| 3 | 20 | 25 | -0.196 |

| $n$ (-) | $K$ (MPa·s$^{\frac{1}{n}}$) | $m$ (-) | $k$ (-) |
|---|---|---|---|
| 2 | 1 | 4 | 3 |

Table 1: Material parameters of amorphous silica (a-SiO$_2$) used in the numerical simulations. $B_0$ and $B_{max}$ are the initial and saturated values of the bulk modulus, $G_0$ and $G_{max}$ are the initial and saturated values of the shear modulus. $P_0$ and $P_1$ are, respectively, the onset and saturation pressures for densification. P', $m$ and $k$ are parameters of the JMAK equation for describing the increase in densification with pressure. $\gamma$ is the saturated value of densification. $n$ and $K$ are viscoplastic parameters in the densification evolution rule.

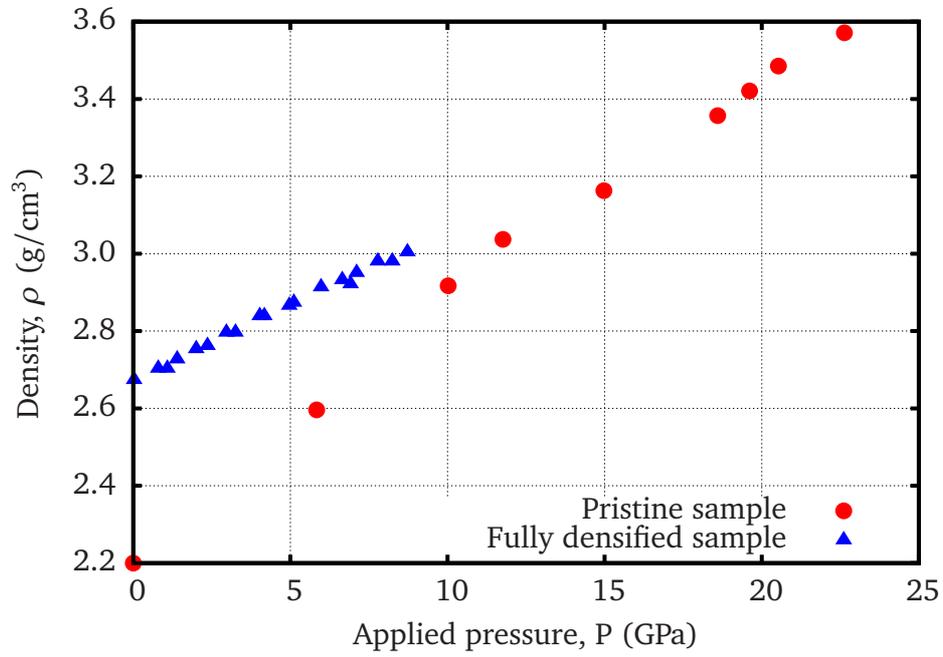

Figure 1: Changes in density during *in situ* testing of silica glass under hydrostatic compression: pristine sample (circles) from Sato and Funamori [9] and fully densified sample (triangles) from Wakabayashi *et al.* [29]

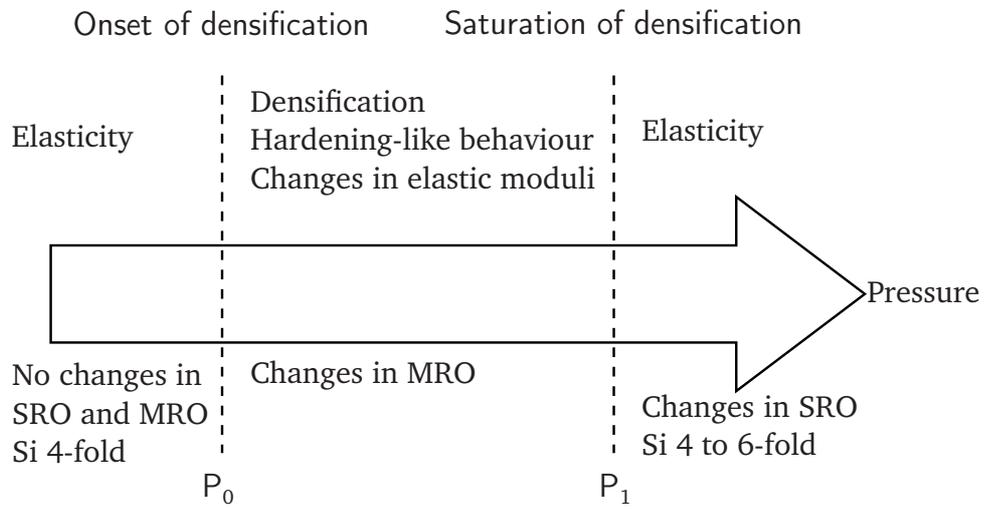

Figure 2: Schematic of the deformation mechanisms in silica glass during hydrostatic compression (above the arrow) alongside the structural changes (below). The arrow stands for the increase in applied pressure. SRO and MRO refer to short range and medium range order, respectively. $P_0$ and $P_1$ are the onset and saturation pressures for densification, respectively.

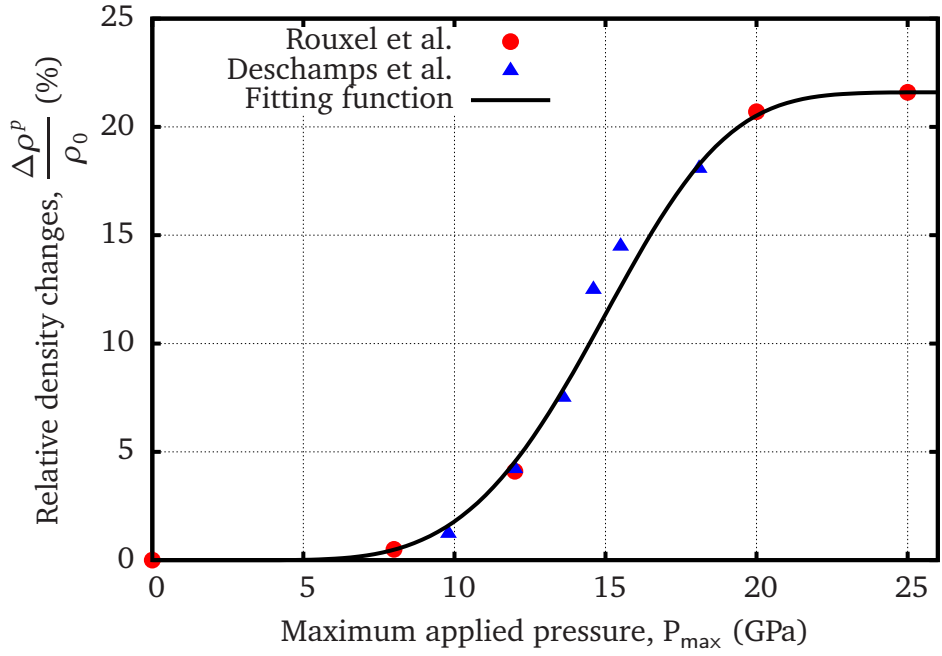

Figure 3: Relative changes in density of silica glass after hydrostatic compression at a maximum applied pressure, from Rouxel *et al.* [19]. A JMAK function is employed to fit the data. It is represented for $P_0 = 3$ GPa and $P' = 20$ GPa here. Data from Deschamps *et al.* [36] are also added.

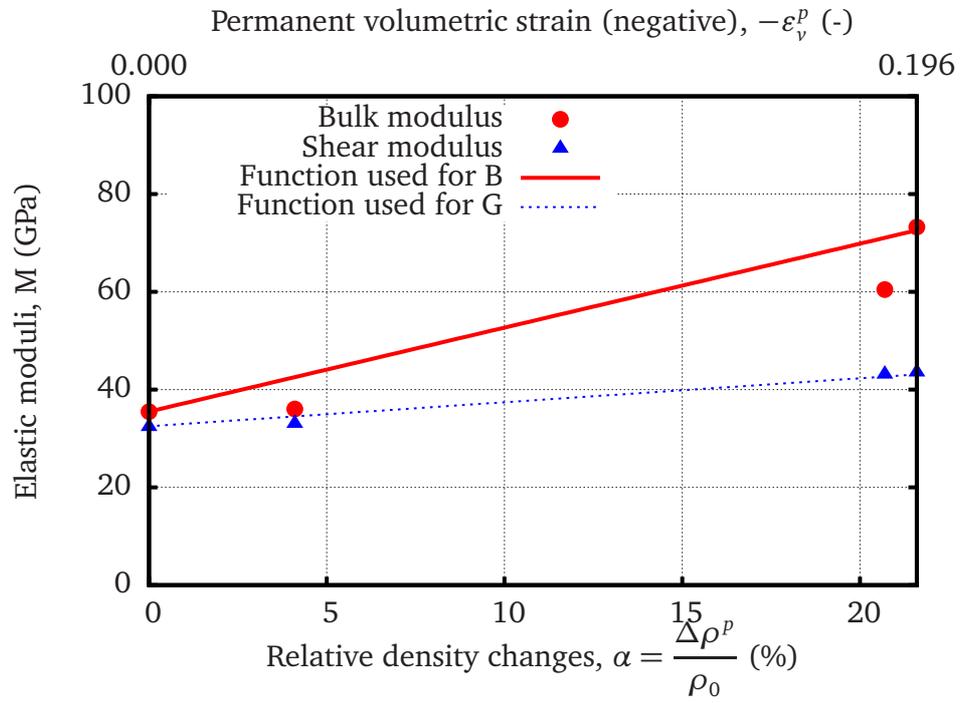

Figure 4: Changes in elastic moduli of silica glass after hydrostatic compression as a function of densification level or permanent volumetric changes, from Rouxel *et al.* [19]. The $\mathscr{C}$ functions (linear) used in the model are indicated.

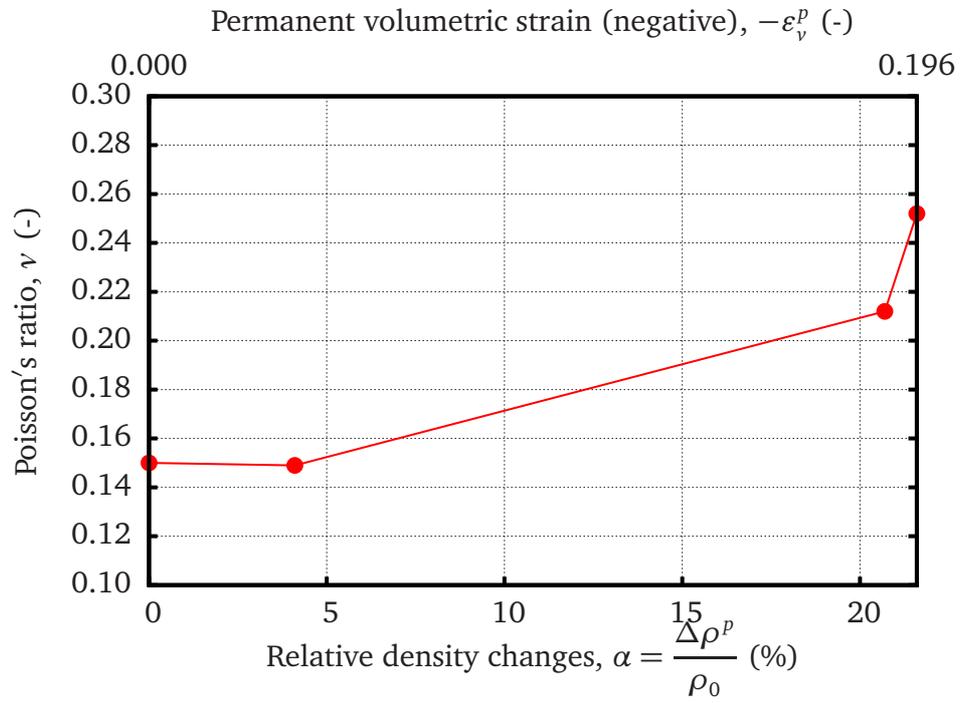

Figure 5: Changes in Poisson's ratio of silica glass after hydrostatic compression at a maximum applied pressure, from Rouxel *et al.* [19]. Lines are guide for the eye.

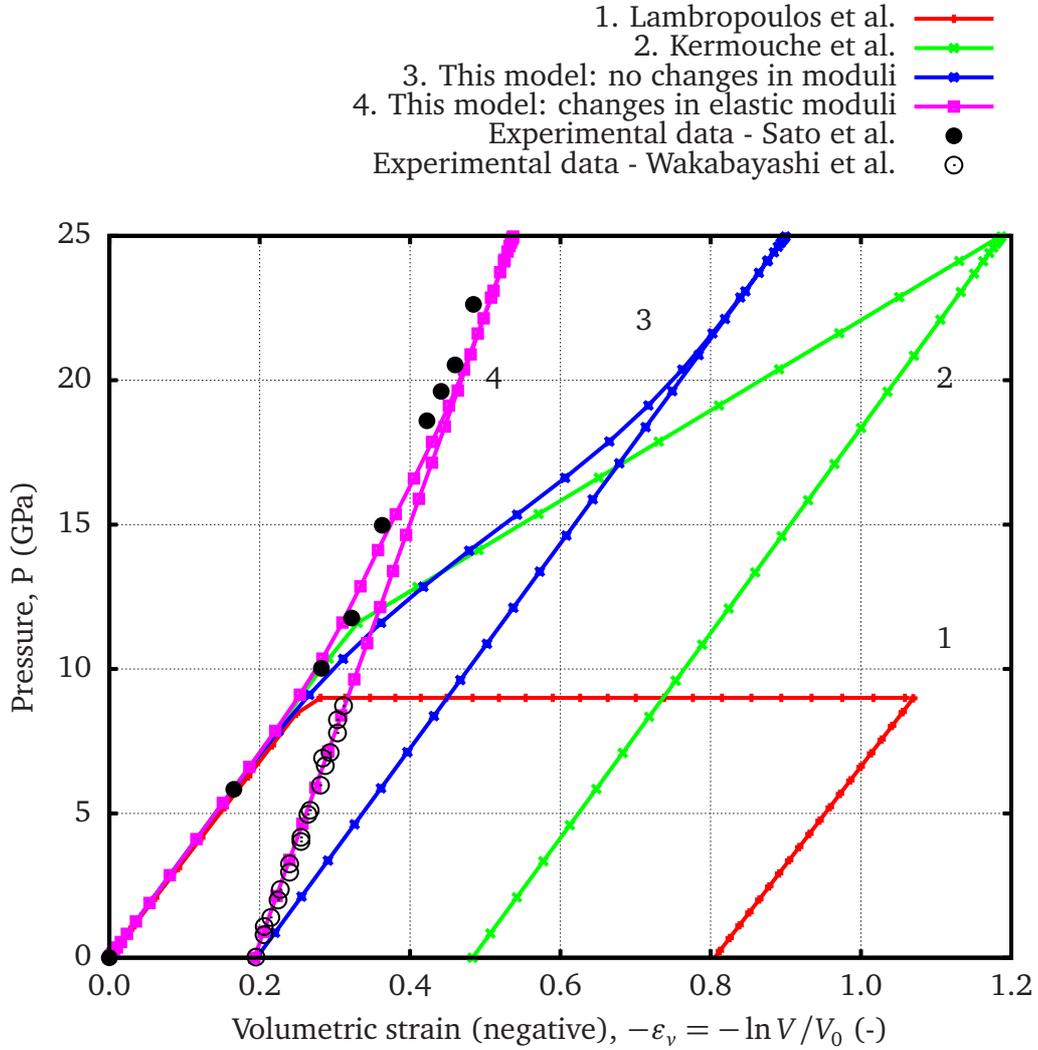

Figure 6: Mechanical response of the hydrostatic compression test on silica glass (pressure vs. volume changes). Experimental data: pristine sample (closed symbols) from Sato and Funamori [9] and fully densified sample (open symbols) from Wakabayashi *et al.* [29]. Comparison with numerical simulations using four different models of growing complexity (labels 1-4).